\documentclass[pre,superscriptaddress,twocolumn]{revtex4-1}
\pdfoutput=1
\usepackage[colorlinks=true,urlcolor=blue]{hyperref}
\usepackage{amsfonts}
\usepackage{graphicx}
\usepackage{mathrsfs}
\usepackage{amsmath}
\usepackage{xcolor}
\usepackage{bm}
\usepackage{titlesec}

\usepackage{color}
\definecolor{red}{rgb}{0.75,0,0}
\definecolor{blue}{rgb}{0,0,0.75}
\definecolor{green}{rgb}{0,0.5,0}

\titlespacing*{\section}
{0pt}{0.5cm}{0.25cm}
\titlespacing*{\subsection}
{0pt}{0.5cm}{0.25cm}

\DeclareMathOperator{\tr}{tr}

\begin{document}

\title{Geometry and Mechanics of Microdomains in Growing Bacterial Colonies}

\author{Zhihong You}
\affiliation{Instituut-Lorentz, Universiteit Leiden, P.O. Box 9506, 2300 RA Leiden, The Netherlands}
\author{Daniel J. G. Pearce}
\affiliation{Instituut-Lorentz, Universiteit Leiden, P.O. Box 9506, 2300 RA Leiden, The Netherlands}
\author{Anupam Sengupta}
\email[Corresponding author:\ ]{anupam.sengupta@uni.lu}
\affiliation{Institute for Environmental Engineering, Department of Civil, Environmental and Geomatic Engineering, ETH Zurich, Stefano-Franscini-Platz 5, 8093 Zurich, Switzerland}
\affiliation{Physics and Materials Science Research Unit, University of Luxembourg, 162 A, Avenue de la Faïencerie, L-1511 Luxembourg City, Luxembourg}
\author{Luca Giomi}
\email[Corresponding author:\ ]{giomi@lorentz.leidenuniv.nl}
\affiliation{Instituut-Lorentz, Universiteit Leiden, P.O. Box 9506, 2300 RA Leiden, The Netherlands}

\begin{abstract}
\vspace*{0.5cm}
Bacterial colonies are abundant on living and nonliving surfaces and are known to mediate a broad range of processes in ecology, medicine and industry. Although extensively researched, from single cells to demographic scales, a comprehensive biomechanical picture, highlighting the cell-to-colony dynamics, is still lacking. Here, using molecular dynamics simulations and continuous modeling, we investigate the geometrical and mechanical properties of a bacterial colony growing on a substrate with a free boundary, and demonstrate that such an expanding colony self-organizes into a ``mosaic'' of microdomains consisting of highly aligned cells. The emergence of microdomains is mediated by two competing forces: the steric forces between neighboring cells, which favor cell alignment, and the extensile stresses due to cell growth that tend to reduce the local orientational order and thereby distort the system. This interplay results in an exponential distribution of the domain areas and sets a characteristic length scale proportional to the square root of the ratio between the system orientational stiffness and the magnitude of the extensile active stress. Our theoretical predictions are finally compared with experiments with freely growing {\em E. coli} microcolonies, finding quantitative agreement. 
\end{abstract}

\maketitle

\section{INTRODUCTION}
Bacteria successfully colonize a plethora of surfaces by producing hydrated extracellular polymeric matrix, generally composed of proteins, exopolysaccharides and extracellular DNA \cite{Mcdougald:2012}. Such surface-associated communities play a crucial role in the pathogenesis of many chronic infections--from benign dental caries in the oral cavity \cite{Rosan:2000, Kaplan:2010} to life-threatening cystic fibrosis and catheter-related endocarditis \cite{Costerton:1999}. In contrast to planktonic populations of motile cells (freely swimming, gliding, or swarming), cells in a sessile colony lack motility. Since most bacteria found in nature exist predominantly as surface-associated colonies \cite{Costerton:1995}, they are permanently exposed to a range of surface-specific forces \cite{Persat:2015}: time-varying internal stress due to growth, contact forces due to interactions with the neighboring cells and substrate they are growing on, or shear stresses due to ambient flows in the system. 

Our understanding of the mechanics of bacterial growth is still in its infancy, specifically in light of the wide range of mechanical cues that single cells overcome to successfully colonize surfaces. Although it has been long known that mechanical forces play a critical role in the development and fitness of eukaryotic cells and, in addition, can regulate key molecular pathways \cite{Hoffman:2011}, the cornerstones of major discoveries in bacterial communities have relied on biochemical pathways triggered exclusively by chemical stimuli \cite{Morris:2008}. Only recently has the role of mechanics in the ecophysiology of prokaryotic cells come to the forefront \cite{Persat:2015, Cho:2007, Volfson:2008, Boyer:2011, Su:2012, Fuentes:2013, Rudge:2013, Farrell:2013, Grant:2014,Sheats:2017}, highlighting the governing biophysical principles that drive colony formation. 

A particularly interesting demonstration of the mechanical aspects of bacterial organization was illustrated in Refs. \cite{Volfson:2008,Boyer:2011,Sheats:2017}, upon confining nonmotile duplicating bacteria in a microchannel. Depending on the channel size, the bacterial population was observed to evolve either into a highly ordered colony \cite{Volfson:2008}, with all the cells parallel to each other and to the channel wall, or, for larger channels, into disordered structures consisting of multiple domains of aligned cells with no global order \cite{Boyer:2011}. More recently, a strikingly similar behavior was identified by Wioland {\em et al}. \cite{Wioland:2016} in suspensions of swimming bacteria. While the existence of an ordered state has been ascribed, in both systems, to a coupling between the bacteria orientation and their collective motion, further enhanced by the geometrical confinement, the origin of the disorder state, as well as its mechanical and statistical properties, is presently unknown.

\begin{figure*}[t]
\centering
\includegraphics[width=1.0\textwidth]{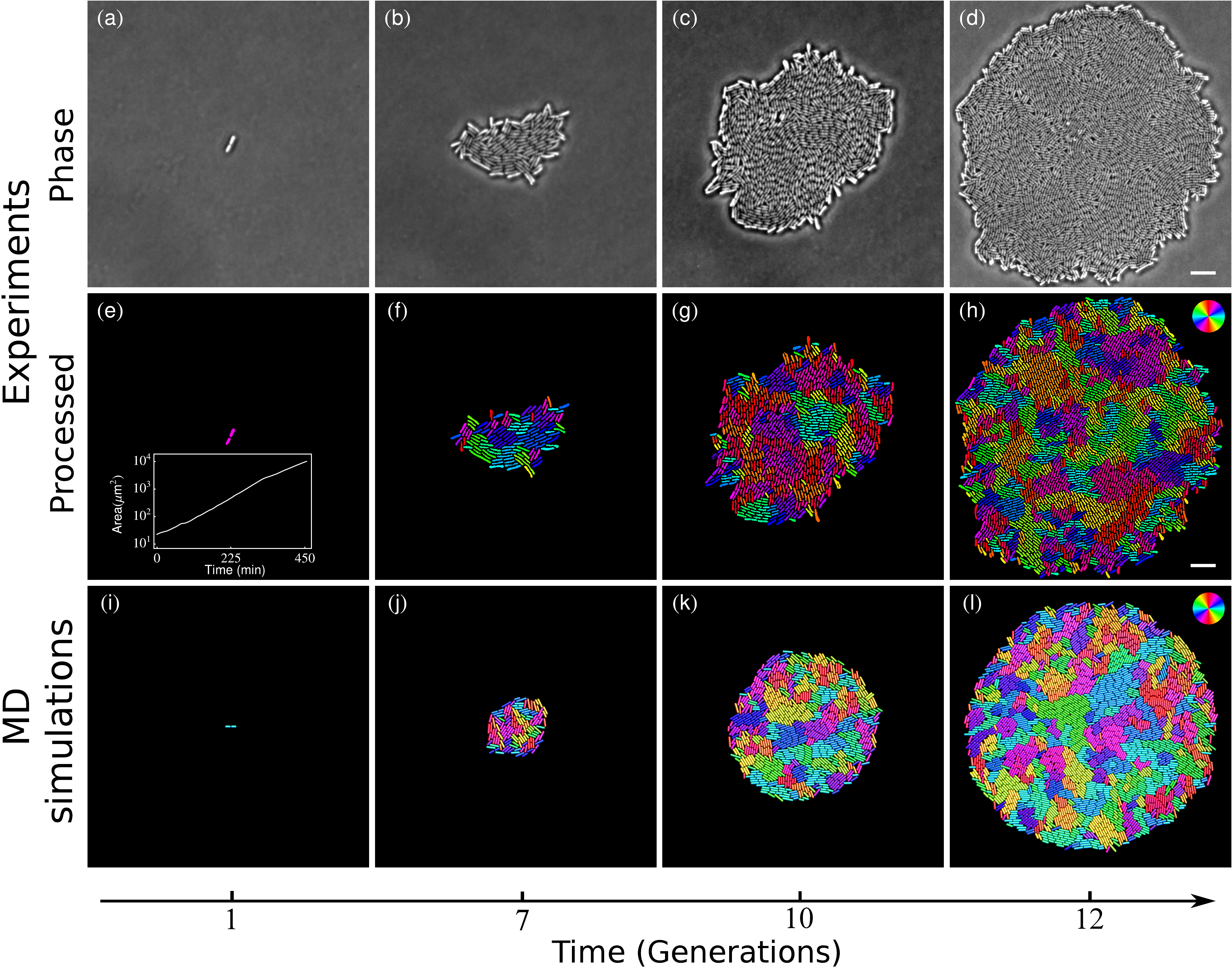}
\caption{\label{fig:ssgrow}{\bf Growth of a bacterial colony.} (a)--(d) Phase-contrast micrographs at different time points capture the growth of a single cell of nonmotile strain of \textit{Escherichia coli} (strain NCM 3722 delta-motA) to a two-dimensional colony under free boundary conditions. The scale bar corresponds to 10 $\mu$m. The cell doubling time was $43.5\pm 2.2$ minutes. After 12 generations (d), the colony was observed to escape into the third dimension and form a second bacterial layer. (e)--(h) Image analyzed snapshots of (a)--(d), capturing the emergence of local orientational order within the growing bacterial colony, represented by differently colored microdomains. Cells are color-coded by the orientation of the domains they belong to, as described in the color wheel in panel (h). The inset in panel (e) plots the area of the growing bacterial colony over time, showing the exponential growth of cells in the colony. (i)--(l) The corresponding time points during the growth of the bacterial colony obtained using molecular dynamics simulations (Sec. \ref{sec:md}). Cells are color-coded with the same method as in panels (e)--(h). By varying the aspect ratio of the cells (length/width) between 1.5 and 4, different physiological states were simulated. 
}
\end{figure*}

In this article, we address this problem and explore the spatial organization and mechanical properties of disordered colonies of sessile bacteria. Our system consists of a colony of nonmotile, rod-shaped bacteria freely growing on a substrate. Although nonmotile bacteria interact typically via steric forces, pushing each other out of the way as they grow in length, the combination of these passive forces with the active bacterial growth results in a complex internal dynamics as well as the emergence of coherent structures [Figs. \ref{fig:ssgrow}(a)--\ref{fig:ssgrow}(d)] reminiscent of those observed in active liquid crystals \cite{Giomi:2015,Sumino:2012,Sanchez:2012,DeCamp:2015,Guillamat:2016}. Using molecular dynamics simulations and continuous modeling, we show that an expanding colony self-organizes into a ``mosaic'' of nematic microdomains, whose typical size is set by a competition between growth-induced active forces that tend to disorder the system and passive steric forces that tend to reorganize the bacteria into closely packed nematic structures. This competition results in an exponential distribution of the domain areas, with a characteristic length scale proportional to the square root of the ratio between the system orientational stiffness and the magnitude of the extensile active stress. Both active and passive forces scale linearly with the cell density. Therefore, despite the colony being denser in the center than at the periphery, such an inherent length scale remains uniform throughout the system. Finally, to assess the significance of our theoretical model, we compare our predictions with experiments on freely growing {\em E. coli} microcolonies (Fig. \ref{fig:ssgrow}). Whereas the statistics of our experiments are not sufficient to make conclusive statements, we do not find obvious discrepancies with our theoretical model. In contrast, the agreement between theory and experiments justifies some degree of optimism and creates promising ground for future experimental research.

This paper is organized as follows: In Sec. \ref{sec:md}, we introduce a hard-rod model for growing bacteria (Sec. \ref{sec:hdmodel}) and describe the geometrical (Sec. \ref{sec:geometry}) and mechanical (Sec. \ref{sec:mechanics}) properties of the emergent microdomains. Building on these results, in Sec. \ref{sec:continuum}, we construct a continuum theory for growing bacterial colonies grounded on the hydrodynamics of active nematic liquid crystals. In Sec. \ref{sec:exp} we present the experimental system and show experimental results in support of our theory. Finally, in Sec. \ref{sec:discussion}, we discuss our results and modeling approach in the context of previously reported experiments and draw conclusions emphasizing the role of geometry and mechanics during the early stages of biofilm formation.

\section{\label{sec:md}HARD-ROD MODEL}
\subsection{\label{sec:hdmodel}Description of the model}

\begin{figure}[t]
\centering
\includegraphics[width=\columnwidth]{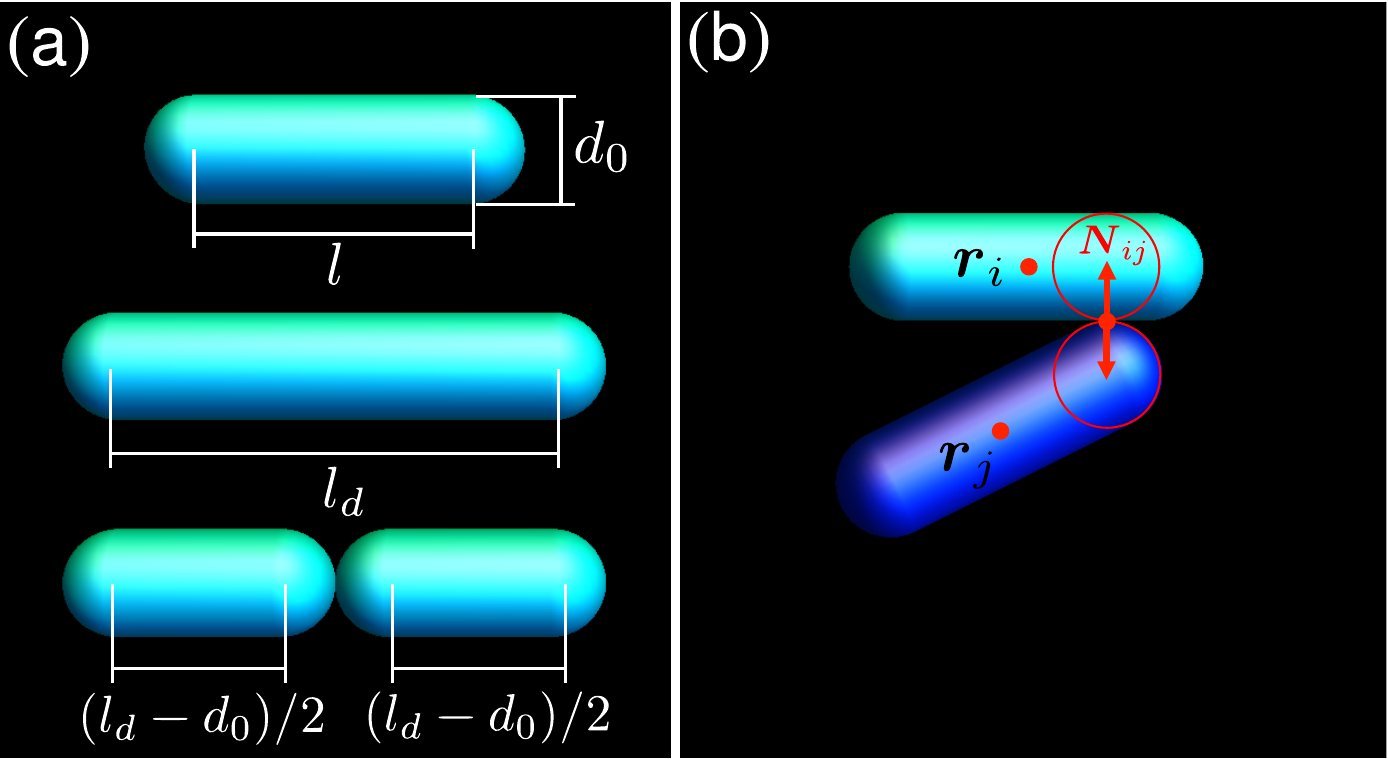}
\caption{\label{fig:smmd} {{\bf Hard-rod model of bacterial growth.} (a) Bacteria are modeled as spherocylinders with a fixed diameter $d_{0}$ and time-dependent length. The cell length $l$ grows linearly in time, and after it reaches the division length $l_{d}$, the cell divides into two identical daughter cells. (b) Cell-cell interaction is modeled via Hertzian forces acting along the normal direction $\bm{N}_{ij}$.}}
\end{figure}

Each bacterium is modeled as a spherocylinder with a fixed diameter $d_{0}$ and a time-dependent length $l$ (excluding the caps on both ends, Fig. \ref{fig:smmd}), growing in a two-dimensional space \cite{Farrell:2013}. The position $\bm{r}_{i}$ and the orientation $\bm{p}_{i}=(\cos\theta_{i},\sin\theta_{i})$ of $i$th cell ($i=1,2,\ldots$), are governed by the over-damped Newton equations for a rigid body \cite{Giomi:2013a}, namely,
\begin{subequations}\label{eq:rods}
\begin{align}
\frac{{\rm d}\bm{r}_{i}}{{\rm d}t} &= \frac{1}{\zeta l_i}\,\sum_{j}\bm{F}_{ij}\;,\\
\frac{{\rm d}\theta_{i}}{{\rm d}t} &= \frac{12}{\zeta l_i^{3}}\,\sum_{j}(\bm{r}_{ij}\times\bm{F}_{ij})\cdot\bm{\hat{z}}\;,	
\end{align}
\end{subequations}
where the summation runs over all the cells in contact with the $i$th cell. The points of contact have positions $\bm{r}_{ij}$ with respect to the center of mass of the $i$th cell and apply Hertzian forces of the form $\bm{F}_{ij}=Y\,d_{0}^{1/2}h_{ij}^{3/2}\bm{N}_{ij}$, where $Y$ is proportional to the Young's modulus, $h_{ij}$ is the overlap distance between the $i$th and $j$th cells, and $\bm{N}_{ij}$ is their common normal unit vector. The constant $\zeta$ is a drag per unit length originating from the substrate adhesion and independent of the cell orientation. The length $l_i$ increases linearly in time, and after it reaches the division length $l_{d}$, the cell divides into two identical daughter cells. In order to avoid synchronization of divisions, the growth rate of each cell, defined as the length increment per unit time, is randomly chosen from an interval $[g/2, 3g/2]$, where $g$ is the average growth rate. Immediately after duplication, the daughter cells have the same orientation as the mother cell but independent growth rates. The rate of cell division can vary over time, with the increase of growth-induced local pressure \cite{Shraiman:2005,Montel:2011}. In bacterial colonies, however, such an effect takes place only at pressure values that are significantly larger than those experienced by the cells in a microcolony \cite{Kumar:2013,Grant:2014} and has, therefore, been neglected in our simulations. This is further demonstrated by the exponential increase of colony area with respect to time, as shown in the inset of Fig. \ref{fig:ssgrow}(e).

Equations \eqref{eq:rods} have been numerically integrated using the following set of parameter values: $d_{0}=1\,\mu$m, $Y=4$ MPa, and $\zeta=200$ Pa h \cite{Farrell:2013}. The division length $l_{d}$ varies from $2\,\mu$m to $5\,\mu$m, and the growth rate varies from $1\,\mu$m/h to $10\,\mu$m/h. The integration is performed with a time step $\Delta t=0.5\times10^{-6}$ h. Each simulation starts with one randomly oriented cell and stops when the total length of the cells in the colony, i.e., $\mathcal{L}=\sum_{i}^{N}(l_{i}+d_{0})$, reaches the value $37500\,d_{0}$, such that colonies with different $l_{d}$ values have approximately the same colony area at the end of the simulation. A typical instance of simulation can be found in Figs. \ref{fig:ssgrow}(i)--\ref{fig:ssgrow}(l). Time-lapse animations showing the growth dynamics of the colonies are included in Ref. \cite{SuppMate:videos}.

In the remainder of the paper, all results are presented in terms of dimensionless quantities, unless otherwise specified. The length is rescaled by the cell diameter $d_{0}$ and the time by $\zeta/Y$. In these units, our hard-rod model has only two free parameters: $l_{d}/d_{0}$, which represents the cell slenderness or aspect ratio, and the rescaled growth rate $g\zeta/(Yd_{0})$.

\subsection{\label{sec:geometry}Stochastic geometry of bacterial colonies}

\begin{figure*}[t]
\centering
\includegraphics[width=\textwidth]{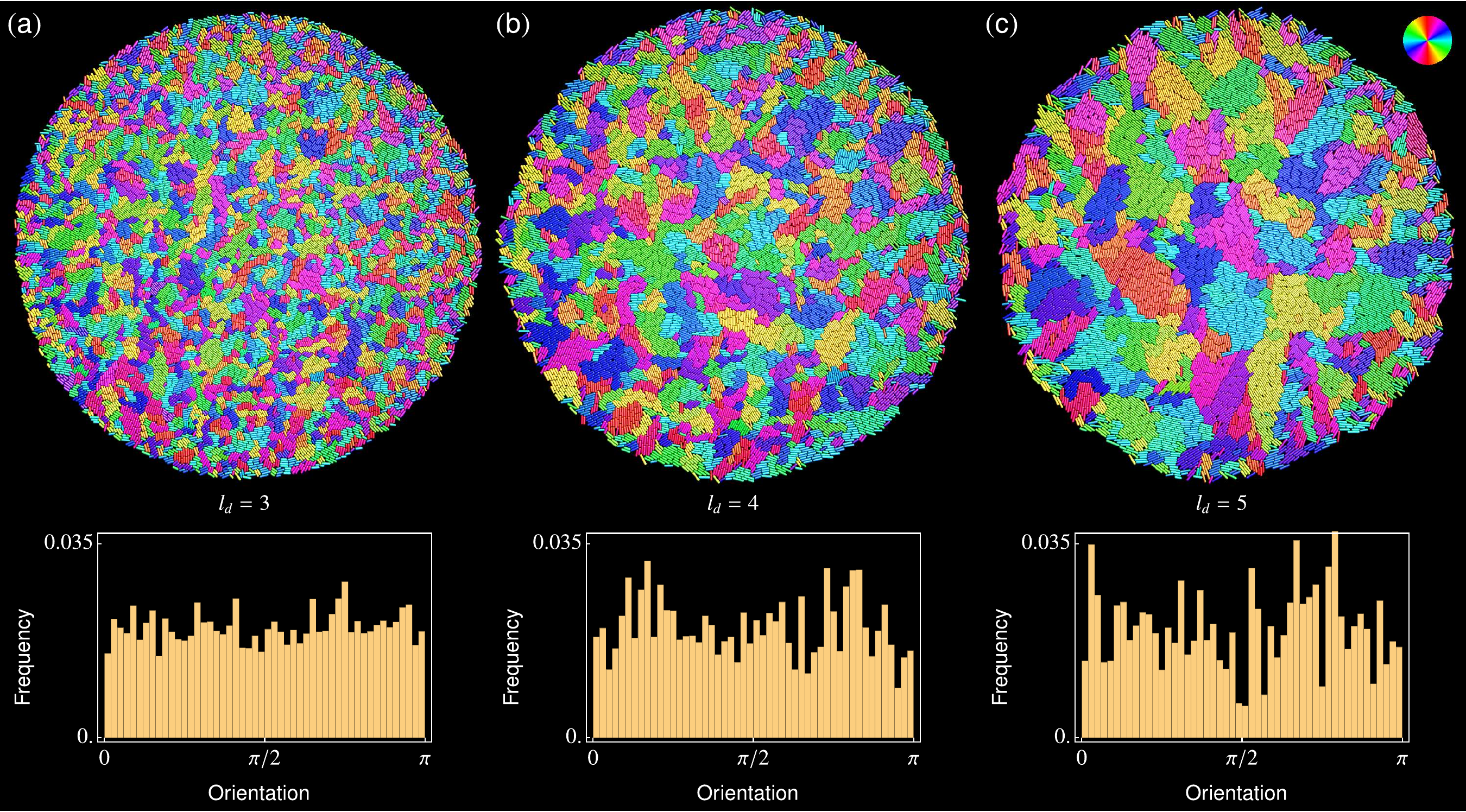}
\caption{\label{fig:ssmd} {\bf Emergence of nematic domains in proliferating bacterial colonies.} (a)--(c) Examples of nematic microdomains in simulated bacterial colonies for various division lengths ($l_d=3,\,4,\,5$, in units of the cell diameter $d_0$). Cells are colored with the same method as in Fig. \ref{fig:ssgrow}. Upon increasing the division length, the typical area of the domains increases progressively. Inside a domain, the cells are highly aligned, while there is no preferential orientation at the scale of the entire colony, as confirmed by the probability distribution of cell orientations (corresponding panels in the lower row).}
\end{figure*}

Figure \ref{fig:ssgrow} shows the typical configurations observed at the early stages of colonization both \textit{in vitro} and \textit{in silico}. The emergence of local nematic order is conspicuous throughout the system; however, this does not propagate across the colony but remains confined to a set of microscopic domains. These nematic domains, or ``patches,'' are separated from each other by fracture lines reminiscent of grain boundaries in crystals \cite{Chaikin:1995,Bowick:2008}. As the colony evolves, the domains grow, merge, buckle, and break apart, in a complex sequence of morphological and topological transformations.

Figure \ref{fig:ssmd} shows three examples of proliferating colonies of cells, each with different $l_{d}$ values and, hence, different cell aspect ratios. The typical domain area, as we can see, increases with the cell aspect ratio. Although the microdomains possess local orientational order, no preferential orientation was observed at the scale of the colony, suggesting that the colony itself is globally isotropic. The absence of the global orientational order can be ascribed to the inherent instability of the domains, which continuously deform and fracture under the effect of growth-induced stress. The typical domain area then represents not only the coherent length scale of orientational order but also the length scale at which the internal stresses compromise. Along the boundary, cells are predominantly tangentially aligned, as a consequence of torque balance. As the forces experienced by the peripheral cells are radial, these cells must orient either tangentially or normally with respect to the boundary in order for the torque acting on them to vanish. Normal alignment is, however, unstable; therefore, most of the peripheral cells are oriented tangentially.

To quantify the emergent geometry of microdomains in a colony, we apply a customized domain segmentation algorithm. Two cells are considered to belong in the same domain if they are in contact, and their relative orientation differed by less than $3\%$. Although decomposition of a colony depends on the chosen threshold, the overall nature of the geometry and the emergent trends identified through different quantifiable parameters are generally robust and independent of the chosen threshold. By using this algorithm, we can then identify domains; measure their positions, orientations, areas \textit{et al.}; and get statistics of these quantities.

\begin{figure}[t]
\includegraphics[width =\columnwidth]{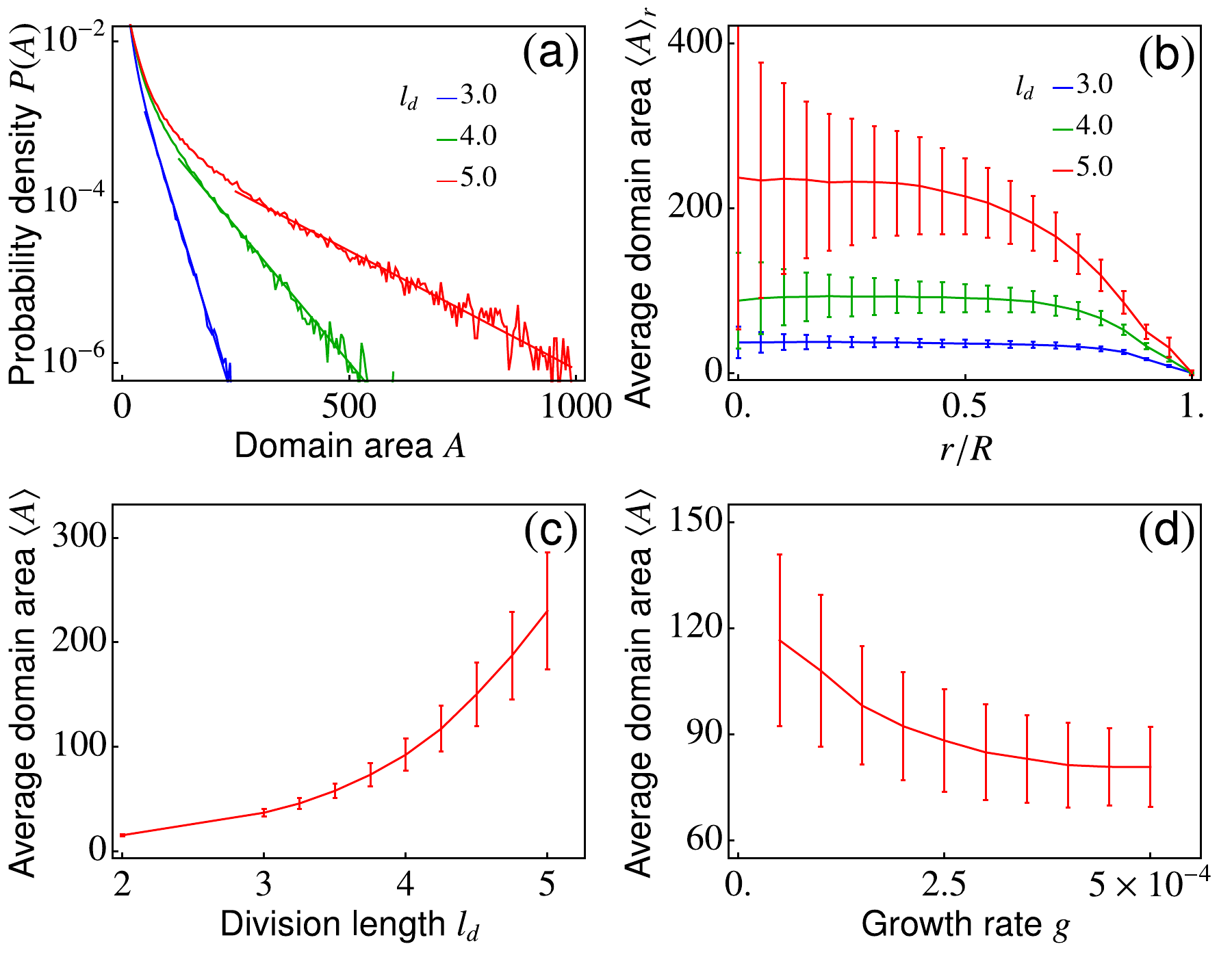}
\caption{\label{fig:geometry}{\bf Geometry of nematic microdomains in bacterial colonies.} (a) Probability distributions of domain area, $P(A)$, for various division length $l_{d}$. The domain area follows the exponential distribution $P(A)\sim \exp(-A/A^{*})$, with $A^{*}=54.3$, $148.2$, and $338.8$, respectively, which increases with the division length $l_{d}$. (b) The average domain area at a distance $r$ from the center of the colony, showing that the area of the domains is found to be constant in the bulk of the colony and drops to zero at the boundary. (c,d) The bulk domain area $\langle A \rangle$ (c) increases with the division length $l_{d}$  and (d) decreases with the growth rate $g$. Here, $\langle A \rangle$ is calculated by averaging the areas of all domains within the range $0 \le r \le R/2$, with $R$ the colony radius. All results shown in panels (a)--(c) correspond to a fixed growth rate of $g=0.0002$ (or $4\mu$m/h in physical units), while panel (d) represents simulation results with a fixed division length $l_{d}=4$.}
\end{figure}

A central quantity to characterize the geometry of a colony is the probability density of the area of these microdomains, $P(A)$. This is shown in Fig. \ref{fig:geometry}(a) for colonies with different $l_{d}$ values. The frequency of domains with area $A$ decreases with $A$ and, for sufficiently large $A$ values, $P(A)$ approaches the exponential distribution:
\begin{equation}
  \label{eq:pdf}
P(A) \sim \exp\left(-\frac{A}{A^*}\right)\;,
\end{equation}
where $A^{*}$ is a characteristic area scale proportional to the average domain area. For small $A$ values, the distribution slightly deviates from the exponential form. This range corresponds to the boundary of the colony where, because of the sudden drop in packing fraction, domains are very small or consist of single cells.

In order to quantify the spatial dependence of the domain, we calculate the average domain area restricted to an annular strip, of width $5d_{0}$, and located at distance $r$ from the center of the colony, i.e., $\langle A \rangle_{r}$ [Fig. \ref{fig:geometry}(b)]. The local domain area is uniform in the bulk of the colony, for a given aspect ratio of cells, before dropping to zero at the boundary, where the colony is more disordered. In turn, the average domain area in the bulk $\langle A \rangle$ is strongly affected by the division length $l_{d}$. This is visibly conspicuous in Fig. \ref{fig:ssmd}. Increasing $l_{d}$ makes the cells, on average, more slender, resulting in larger and more stable domains, as revealed by the plot in Fig. \ref{fig:geometry}(c). More interestingly, increasing the growth rate $g$ has the opposite effect and causes a drop in the domain area [Fig. \ref{fig:geometry}(d)]. All data in Fig. \ref{fig:geometry}, as well as those in Fig. \ref{fig:mechanics}, are obtained by averaging over $480$ runs. All results are obtained by analyzing the configurations of the colonies at which the simulations stop, unless otherwise specified. The error bars in the two figures show the standard deviations of the $480$ samples with respect to the mean values.

The results reported in this section quantitatively demonstrate that the spatial organization of the microdomains in expanding bacterial colonies is regulated by the competing effects of the cell aspect ratio and the growth rate. These effects can ultimately be ascribed to the mechanical properties of the system, as we explain in the next subsection. We stress here that our approach does not aim to faithfully reproduce all the experimental details but rather to provide a conceptual key for understanding certain geometrical and mechanical properties, with the help of a minimal model comprising a single fitting parameter: i.e., the timescale $\tau=\zeta/Y$. Other properties, such as the roughness of the colony edge and smoother variation in the orientation of neighboring domains, are not well captured by our simple model and would require a more sophisticated construction, accounting for the adhesive interaction between neighboring cells, the flexibility of the cell membrane, and more specific cell-substrate interactions \cite{Duvernoy:2018}. Unfortunately, this would imply a cost in terms of free parameters and reduced simplicity in the interpretation of the numerical results.

\subsection{\label{sec:mechanics}Mechanical properties}

\begin{figure*}[t]
\includegraphics[width = \textwidth]{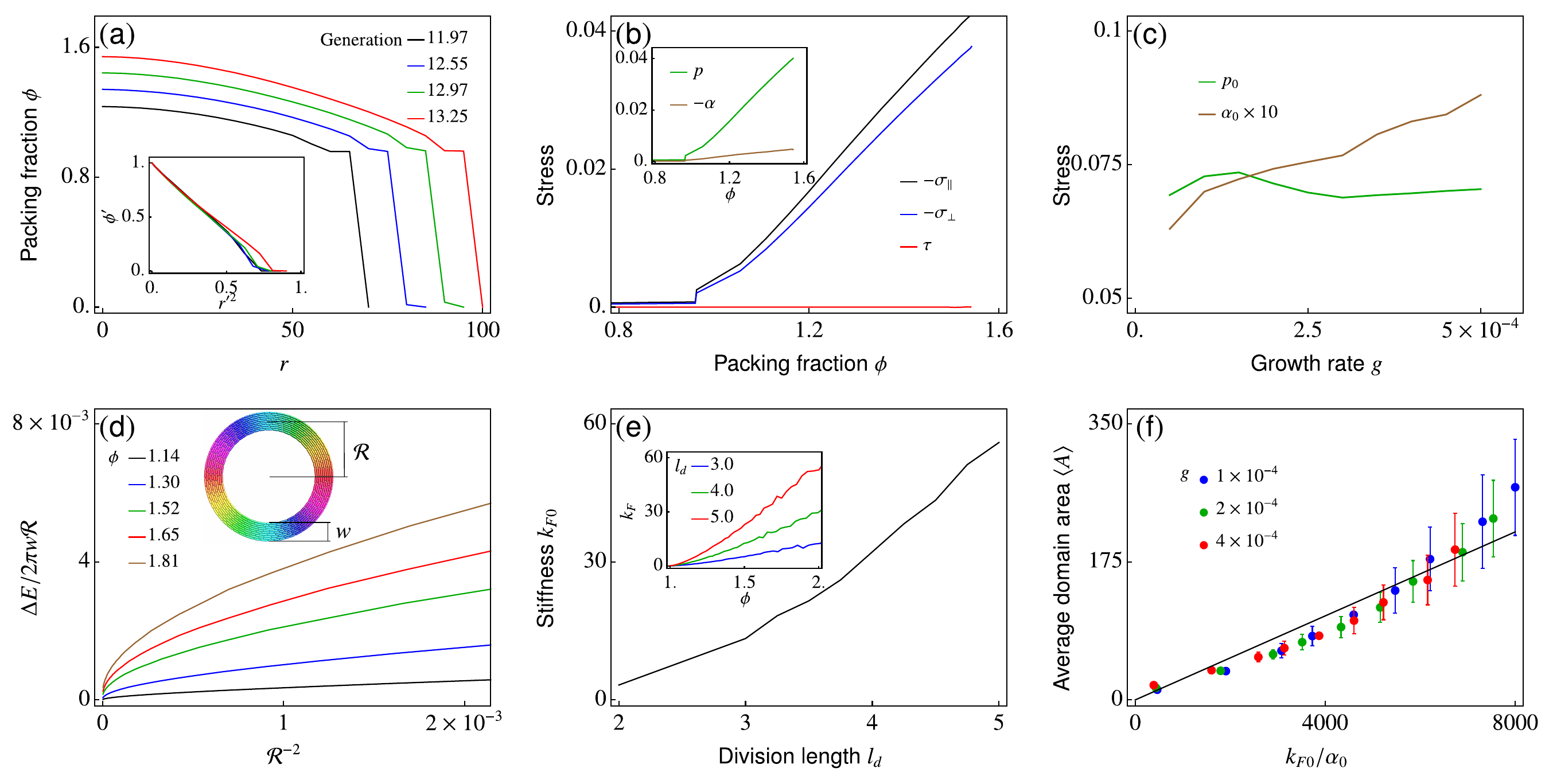}
\caption{\label{fig:mechanics}{\bf Mechanical properties of microdomains in growing bacterial colonies.} (a) Spatial dependence of the packing fraction for different ages of colonies. The inset shows $\phi'=[\phi(r)-\phi(R)]/[\phi(0)-\phi(R)]$ versus $r'^{2}=(r/R)^{2}$. All the curves collapse on the same line as demanded by Eq. \eqref{eq:density_profile}. (b) Different components of the internal stress $\bm{\sigma}$ as functions of packing fraction $\phi$. The normal stress parallel to the director $\bm{n}$, $|\sigma_{\parallel}|$, is larger than that perpendicular to it, i.e., $|\sigma_{\perp}|$. Both $|\sigma_{\parallel}|$ and $|\sigma_{\perp}|$ are piecewise linear functions of $\phi$, while the shear component $\tau$ vanishes. The normal components of stress can be rearranged into a hydrostatic pressure $p$ and an extensile active stress $\alpha$, and both increase linearly with the packing fraction (inset). In both panels (a) and (b), the growth rate and the division length are fixed, i.e., $g=0.0002$ and $l_{d}=4$. Both $\phi$ and $\boldsymbol{\sigma}$ are averaged over a thin annulus of radius $r$ and width $5d_{0}$, centered at the colony center. (c) The pressure is independent of growth rate $g$, while the active stress increases with it. Here, $l_{d}=4$ is fixed. (d) Difference of energy density(energy per unit area) between the straight channel and a ring-shaped channel of radius $\mathcal{R}$, as a function of $\mathcal{R}^{-2}$. An example of a ring-shaped channel is shown in the inset, with radius $\mathcal{R}$ and width w. Cells are colored by their orientations according to the color wheel in Fig. \ref{fig:ssgrow}. (e) The orientational stiffness $k_{F}$ increases linearly with the packing fraction (inset), and the prefactor of the linear fit, $k_{F0}$, increases with the division length $l_d$. (f) The average domain area $\left<A\right>$ is approximately proportional to $k_{F0}/\alpha_0$, for various combinations of growth rate and division length. We choose three growth rates (identified by colors), and for each growth rate, we gradually increase the division length from $l_d=2$ to $l_d=5$, corresponding to different data points with the same color. In all results presented, the length is expressed in units of the cell width, $d_{0}$, and time in units of the timescale $\zeta/Y$ defined in Eqs. \eqref{eq:rods}.}
\end{figure*}

The domain geometry in a proliferating bacterial colony is determined by the interplay between two competing forces: steric repulsion between neighboring cells and the extensile stresses due to cell growth. While cell-cell steric repulsion favors alignment, the emergent extensile stresses due to the growth within a restricted environment (i.e., the space delimited by the neighboring domains) tend to deform and eventually fracture a domain. Both of these effects are due to contact forces and are, therefore, enhanced by the local packing fraction $\phi$. To clarify this concept, we measure the local packing fraction $\phi(r,t)=\sum_i a_{i}(t)/
\mathcal{A}_r$, where $a_{i}(t)$ is the area of the $i$th cell, located at time $t$ inside a thin annulus of radius $r$, width $5d_{0}$, and area $\mathcal{A}_{r}$, centered at the colony center. We find that the colony has a radial symmetry; hence, the local packing fraction depends exclusively on the distance $r$ from the center. Figure \ref{fig:mechanics}(a) shows that at any given time, the packing fraction decreases monotonically with $r$. As bacteria duplicate and progressively colonize the surrounding space, the local packing fraction increases with time throughout the system while maintaining a characteristic spatial profile that smoothly interpolates between a time-dependent maximum $\phi(0,t)=\phi_{\max}(t)$, at the center of the colony, and a time-independent minimum, $\phi(R,t) = \phi_{c}$, at the edge ($R$ being the colony radius). The quantity $\phi_{c}\approx 1$ is the critical packing fraction at which the cells first start to compress each other. In close proximity of the edge of the colony, $\phi <\phi_{c}$, and the contact forces tend to reorient the cells without compressing them, leading to an abrupt drop in the packing fraction. Upon rescaling the packing fraction by $\phi(0)-\phi(R)$ and the distance $r$ by the colony radius $R$, the spatial dependence of the pacing fraction can be described, at any time, by a simple quadratic law:
\begin{equation}\label{eq:density_profile}
\frac{\phi(r)-\phi(R)}{\phi(0)-\phi(R)} = 1-\left(\frac{r}{R}\right)^{2}\;,
\end{equation}
as illustrated in the inset of Fig. \ref{fig:mechanics}(a). As we analytically prove in Sec. \ref{sec:continuum}, such a density profile originates from the balance between growth-induced pressure and drag from the substrate.

The tendency of the cells to align with each other is driven by the local steric interactions and can be conceptualized in the framework of Frank elasticity \cite{Landau:1986}, starting from the free-energy density:
\begin{equation}\label{eq:frank_energy}
f_{F} = \frac{1}{2} k_{F} |\nabla\bm{n}|^{2}\;.	
\end{equation}
Here, $k_{F}$ is an orientational stiffness penalizing, in equal amounts, splay and bending deformations, and $\bm{n}$ is the local nematic director corresponding to the average orientation of the bacteria in a local region. Any departure from the uniformly aligned configuration causes restoring forces proportional to the field $\bm{h}=-\delta / \delta \bm{n}\int {\rm d}A\,f_{F} = k_{F}\nabla^{2}\bm{n}$ \cite{Landau:1986}. As a consequence of growth, each cell further acts as an extensile force dipole that pushes away its neighbors along the $\pm\bm{n}$ direction. This collectively gives rise to an internal stress of the form
\begin{equation}\label{eq:active_stress}
\bm{\sigma} = -p \bm{I}+\alpha\left(\bm{n}\bm{n}-\frac{1}{2}\,\bm{I}\right)\;,
\end{equation}
where $p$ is the pressure, $\bm{I}$ the identity matrix, and $\alpha$ the deviatoric active stress \cite{Pedley:1992,Hatwalne:2004}. In the most general case, the three quantities $k_{F}$, $p$, and $\alpha$, appearing in Eqs. \eqref{eq:frank_energy} and \eqref{eq:active_stress}, are functions of the local packing fraction and the nematic order parameter, in addition to the cell aspect ratio and the growth rate. 

Equations \eqref{eq:frank_energy} and \eqref{eq:active_stress} identify a fundamental length scale $\ell_{\rm a}=\sqrt{k_F/|\alpha|}$, proportional to the distance at which the passive restoring forces arising in the system, in response to a local distortion, balance the active forces that cause the nematic director to rotate \cite{Giomi:2015}. This length scale plays a pivotal role in the mechanics of active nematic liquid crystals \cite{Voituriez:2005,Marenduzzo:2007,Edwards:2009,Giomi:2011,Giomi:2012} and, as we clarify later, determines their collective behavior and mechanical properties. In the following, we demonstrate that, in a growing colony of nonmotile cells, the inherent length scale $\ell_{\rm a}$ determines the geometrical properties of the microdomains in such a way that $\langle A \rangle \sim \ell_{\rm a}^{2}$. For this purpose, we measure the orientational stiffness $k_{F}$ and the stresses $\bm{\sigma}$ exerted inside the colony. The stress experienced by the $i$th cell, $\boldsymbol{\sigma}_{i}$, can be calculated from the virial expansion \cite{Volfson:2008}:
\begin{equation}\label{eq:virial_stress}
\bm{\sigma}_{i}=\frac{1}{2a'_{i}}\sum_{j} \bm{r}_{ij}\,\bm{F}_{ij}\;,
\end{equation}
where $a'_i=a_i/\phi$ is the effective area occupied by the $i$th cell. We express the tensor in the basis of the nematic director and its normal $\bm{n}^{\perp}=(-n_{y},n_{x})$, namely,
\begin{equation}\label{eq:stress}
\bm{\sigma} = \sigma_{\parallel}\bm{n}\bm{n}+\sigma_{\perp}\bm{n}^{\perp}\bm{n}^{\perp}+\tau(\bm{n}\bm{n}^{\perp}+\bm{n}^{\perp}\bm{n})\;.	 
\end{equation}
Figure \ref{fig:mechanics}(b) shows a plot of the various components of the stress tensor versus the packing fraction, given by Eq. \eqref{eq:virial_stress}. As expected, the normal stresses $\sigma_{\parallel}$ and $\sigma_{\perp}$ increase with the packing fraction and, at any finite packing fraction, are such that $|\sigma_{\parallel}|>|\sigma_{\perp}|$, as a consequence of the anisotropic cell growth. The shear stress $\tau$, on the other hand, is always negligible because of the absence of lateral friction between the cells. Note that both $\sigma_{\parallel}$ and $\sigma_{\perp}$ are negative because of the extensile nature of the growth-induced forces. The dependence of the normal stresses on the packing fraction is piecewise linear: For $\phi<\phi_{c}$, the contact forces can be relieved by rotations and repositioning of the cells, and $\sigma_{\parallel} \approx \sigma_{\perp} \approx 0$; however, for $\phi>\phi_{c}$, the cells in the bulk are tightly packed, and internal stresses build up as the packing fraction increases.  Setting $\tau=0$ in Eq. \eqref{eq:stress} and taking $\bm{n}^{\perp}\bm{n}^{\perp}=\bm{I}-\bm{n}\bm{n}$, one can rearrange the stress tensor in the form
\begin{equation}
\bm{\sigma} = -\frac{|\sigma_{\parallel}+\sigma_{\perp}|}{2}\,\bm{I}+(\sigma_{\parallel}-\sigma_{\perp})\left(\bm{n}\bm{n}-\frac{1}{2}\,\bm{I}\right)\;.
\end{equation}
Comparing this with Eq. \eqref{eq:active_stress} straightforwardly yields $p=(|\sigma_{\parallel}+\sigma_{\perp}|)/2$ and $\alpha=\sigma_{\parallel}-\sigma_{\perp}$. Together with the numerical results summarized in Fig. \ref{fig:mechanics}, this implies
\begin{equation}\label{eq:stress_scaling}
p=p_0 (\phi-\phi_{c})\;,\qquad
\alpha= -\alpha_0 |\phi-\phi_{c}|\;,
\end{equation}
as long as $\phi>\phi_{c}$. Not unexpectedly, the longitudinal growth of the cells gives rise to an extensile (i.e., $\alpha<0$) active stress that decreases monotonically with the distance from the center of the colony. The prefactors $p_{0}$ and $\alpha_0$ are plotted in Fig. \ref{fig:mechanics}(c) as a function of the growth rate $g$. The active stress $\alpha_{0}$ increases monotonically with $g$, while $p_0$ is essentially independent. 

In order to estimate the orientational stiffness $k_{F}$, we place our {\em in silico} bacterial colony inside an annular channel of width $w=10d_{0}$ and radius $\mathcal{R}$ [Fig. \ref{fig:mechanics}(d), $w\ll \mathcal{R}$], and calculate the energy associated with the Hertzian contacts: $E=(2/5)\,Y d_{0}^{1/2}\sum_{\langle ij \rangle}h_{ij}^{5/2}$, where the summation runs over all the pairs of cells in contact with each other. By comparing how the energy density changes with the curvature of the channel, we can infer the orientational stiffness. Figure \ref{fig:mechanics}(d) shows a plot of the difference $\Delta E = E(\phi,\mathcal{R})-E(\phi,\infty)$ between the energy of a bent channel with radius $\mathcal{R}$ and a straight channel (both have a length $2\pi \mathcal{R}$), normalized by the area $2
\pi w \mathcal{R}$ of the channel, as a function of the squared curvature $\kappa^{2}=1/\mathcal{R}^{2}$. From Eq. \eqref{eq:frank_energy}, it follows that $k_{F}=\partial_{\kappa^{2}}\Delta E/(2\pi 
 w \mathcal{R})|_{\kappa=0}$. As shown in the inset of Fig.~\ref{fig:mechanics}(e), the orientational stiffness $k_{F}$ increases linearly with the packing fraction, i.e., $k_F=k_{F0} (\phi-\phi_{c})$. Furthermore, increasing the slenderness of the cells makes the colony orientationally stiffer [Fig.~\ref{fig:mechanics}(e)].

Combining the measurements of the extensile active stress and the orientational stiffness, we are finally able to formulate a scaling law for the area of the nematic microdomains comprising our simulated bacterial colonies. Namely,
\begin{equation}\label{eq:average_area}
\langle A \rangle \sim \frac{k_F}{|\alpha|}\;,
\end{equation}
in agreement with our numerical data [Fig.~\ref{fig:mechanics}(f)]. In summary, bacterial colonies freely growing on a two-dimensional frictional substrate spontaneously organize into a ``mosaic'' of microdomains consisting of highly aligned cells. The domains are randomly oriented so that the colony is globally isotropic and circularly symmetric at the global scale, while their areas are exponentially distributed, as indicated in Eq. \eqref{eq:pdf}. Such a distribution results from the competition between passive steric forces, which favor local alignment, and the extensile active forces originating from the cell growth. These forces balance at the length $\ell_{\rm a}=\sqrt{k_{F}/|\alpha|}$, resulting in a characteristic domain area that scales as $\ell_{\rm a}^{2}$. Remarkably, both the orientational stiffness $k_{F}$ and the extensile active stress $\alpha$ scale linearly with the packing fraction $\phi$. Consequently, $k_{F}/\alpha=k_{F0}/\alpha_{0}$, so the average domain area is uniform throughout the colony [Fig. \ref{fig:geometry}(b)]. Such a cancellation of the dependence is intriguing: It is presumably specific to the type of interactions chosen here, which we do not expect to hold in general. Including the bending elasticity of the cells could, for instance, change the packing fraction dependence of $k_{F}$ and $\alpha$, resulting in a space-dependent active length scale. Yet, the mechanism described here and summarized by Eq. \eqref{eq:average_area} is general and does not depend on the details of the model.

\section{\label{sec:continuum}CONTINUOUS MODEL}

In this section, we demonstrate that much of the behavior previously described can be quantitatively captured in the realm of continuum mechanics by means of a suitable extension to the hydrodynamic equations of active nematic liquid crystals. These have been successfully used in the past decade to describe a variety of active fluids, typically of biological origin, consisting of self-propelled or mutually propelled apolar building blocks, such as \textit{in vitro} suspensions of microtubules and kinesin \cite{Ahmadi:2005,Ahmadi:2006,Liverpool:2007,Giomi:2013b,Giomi:2014,Giomi:2015,Thampi:2013,Doostmohammadi:2016a,Doostmohammadi:2017}, microswimmers \cite{Lau:2009}, and cellular monolayers \cite{Doostmohammadi:2015}. Recently, attempts have been made to describe sessile bacteria, in the language of nematic liquid crystals \cite{Volfson:2008,Doostmohammadi:2016a}. Here, we introduce a comprehensive hydrodynamic framework, incorporating the density effects described in the previous section as well as the deviatoric active stresses in the colonization dynamics. 

\begin{figure}[t]
\includegraphics[width=1.0\columnwidth]{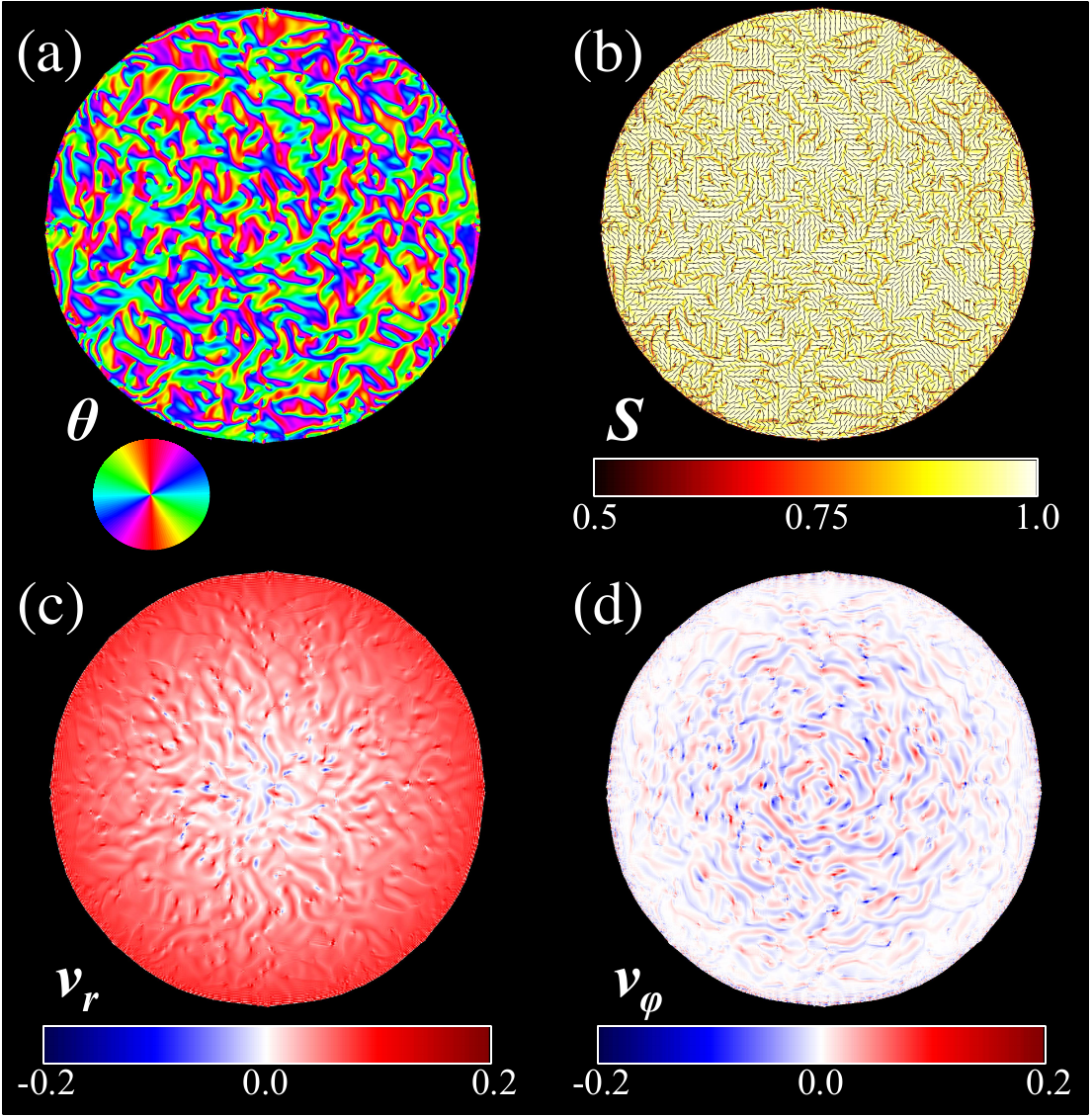}
\caption{\label{fig:sshd}{\bf Continuous Model.} (a) Snapshot of a typical configuration obtained from a numerical integration of Eqs. \eqref{eq:hydrodynamic}. Displayed here is the angle between the nematic director and the $x$ axis, colored using the same color scheme as in Fig.~\ref{fig:ssgrow}. (b) Director field (lines) superimposed on a color map of the nematic order parameter $S$. As for the hard-rod model, nematic order is approximatively uniform except at the boundary of the domains. (c) Radial and (d) tangent components of the velocity field, $v_{r}$ and $v_{\varphi}$. Along the radial direction, the flow is predominantly expansive because of the cell growth. On the other hand, there is no net circulation along the tangential direction.}
\end{figure}

An expanding bacterial colony can be described in terms of the material fields $\rho$, $\bm{v}$, and $\bm{Q}$, representing, respectively, the cell density, velocity, and the nematic order. The latter is represented via the two-dimensional tensor field $\bm{Q} = S(\bm{n}\bm{n}-\bm{I}/2)$ \cite{DeGennes:1993}, where $0 \le S \le 1$ is an order parameter quantifying the local nematic order of the cells. The dynamics of these fields is then governed by the following hydrodynamic equations \cite{Giomi:2012}:
\begin{subequations}\label{eq:hydrodynamic}
\begin{align}
&\frac{D\rho}{Dt} = k_{g}\rho + \mathcal{D}\nabla^2\rho\;,\\[5pt]
&\frac{D(\rho\bm{v})}{Dt} = \nabla\cdot\bm{\sigma}-\xi\rho\bm{v}\;,\\[5pt]
&\frac{D\bm{Q}}{Dt} = \lambda S \bm{u}+\bm{Q}\cdot\bm{\omega}-\bm{\omega}\cdot\bm{Q}+\gamma^{-1}\bm{H}\;,
\end{align}
\end{subequations}
where $D/Dt=\partial_{t}+\bm{v}\cdot\nabla+(\nabla\cdot\bm{v})$ is the material derivative. Equation (\ref{eq:hydrodynamic}a) describes an exponential growth of the colony total mass at rate $k_{g}$ (proportional to the length extension rate $g$ used in Sec. \ref{sec:md}). As the cells duplicate, they are transported across the colony by convective currents. An additional diffusive term, with $\mathcal{D}$ a small diffusion coefficient, is introduced for regularization. The cells' momentum density $\rho\bm{v}$ is subject to the internal stresses $\bm{\sigma}$ as well as the frictional force $-\xi\rho\bm{v}$. The former can, in turn, be expressed as
\begin{equation}\label{eq:continuum_stress}
\bm{\sigma} = -p\bm{I} + \alpha \bm{Q} - \lambda S\bm{H} + \bm{Q}\cdot\bm{H}-\bm{H}\cdot\bm{Q}\;,	
\end{equation}
where the first three terms represent, as in Eq. \eqref{eq:active_stress}, the isotropic pressure and extensile active stress introduced by the cell growth. The remaining terms describe the elastic stresses in the colony arising from the passive aligning interactions between the cells. The tensor field $\bm{H}$ in Eqs. (\ref{eq:hydrodynamic}c) and \eqref{eq:continuum_stress} can be defined starting from the Landau--de Gennes free-energy density:
\begin{equation}\label{eq:landau_de_gennes}
f_{\rm LdG} = \frac{1}{2}L_{1}|\nabla\bm{Q}|^{2} + \frac{1}{2}A_{2}\tr{\bm{Q}^{2}}+\frac{1}{4}A_{4}\left(\tr{\bm{Q}^{2}}\right)^{2}\;,
\end{equation}
as $\bm{H}=-\delta/\delta\bm{Q}\int {\rm d}A\,f_{\rm LdG}$. Here, $L_{1} \sim k_{F}$ is an orientational stiffness, while the functions $A_{2}$ and $A_{4}$ set the boundary between the isotropic ($S=0$) and nematic ($S>0$) phases. At equilibrium, $\bm{H}=\bm{0}$ and $S=\sqrt{-2A_{2}/A_{4}}$. In our system of growing cells, orientational order is driven uniquely by the steric repulsion, and the system transitions to a nematic phase for large enough densities. We set $A_{2}=A_0 (\rho^{*}-\rho)/2$ and $A_{4} =A_0 \rho$, so the system has an equilibrium order parameter $S=\sqrt{1-\rho^{*}/\rho}$, with a critical density $\rho^*$; hence, the colony is disordered for densities $\rho < \rho^*$, and it is nematic for $\rho>\rho^*$. Finally, consistently with Eq. (\ref{eq:landau_de_gennes}), the nematic tensor relaxes toward the minimum of the Landau--de Gennes energy, with $\gamma$ a rotational viscosity, while rotating as a consequence of the internal motion of the cells. This effect is embodied in the first two terms of Eq. (\ref{eq:hydrodynamic}c), with $u_{ij}=(\partial_{i} v_{j}+\partial_{j} v_{i}-\delta_{ij}\nabla\cdot\bm{v})/2$ and $\omega_{ij}=(\partial_{i} v_{j}-\partial_{j} v_{i})/2$ representing strain rate and the vorticity tensor, respectively, and $\lambda$ the flow-alignment parameter \cite{Giomi:2012}.

\begin{figure}[t]
\includegraphics[width=1.0\columnwidth]{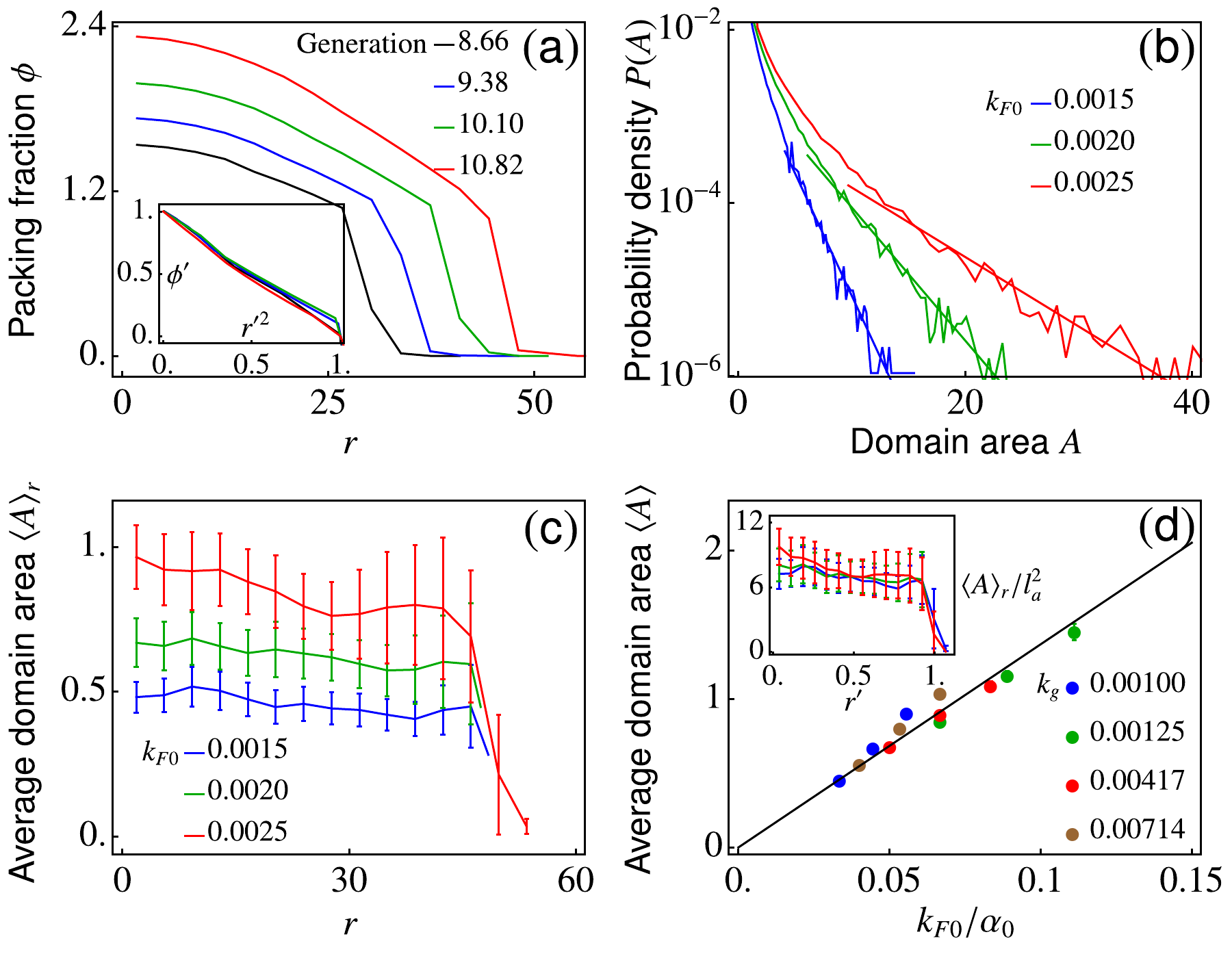}
\caption{\label{fig:resuhd}{\bf Geometrical properties of continuous colonies.} (a) Spatial distribution of the packing fraction ($\phi = \rho/\rho_c$) for colonies of different ages, given by their generation. Similar to the hard-rod model, the inset shows that all curves collapse to a single line when $\phi$ and $r$ are rescaled by $\phi'=[\phi(r)-\phi(R)]/[\phi(0)-\phi(R)]$ and $r'=r/R$. (b) Probability density of the domain area, $P(A)$, for various values of the orientational stiffness $k_{F0}$. (c) The average domain area at a distance $r$ from the center of the colony. As in the hard-rod model, the typical domain area is uniform across the colony. (d) The average domain area for a colony scales linearly with the squared active length scale $\ell_{\rm a}^{2}=k_{F}/|\alpha|$. The inset shows that the radial distribution of domain areas can be rescaled by the squared active length scale to the same value. In presenting the results, we use $l = 1/\sqrt{\rho_c}$ as our units of length. We interpret the doubling time $t_g=\log(2)/k_g$ as the time per generation. The simulations were run on a $351\times 351$ grid with the spacing set to $1$. They start from a circle of bacteria with density $\rho_0=0.1$ and radius $6$ grid points and grow to a given total mass, at which point the simulation ends. The (unscaled) parameters used were $\rho^*=0.005$, $A_0=50$, $\rho_c=0.1$, $\xi=5$, $P_0=10$, $\lambda=0.1$, and $\zeta=10$. The two variable parameters are $\alpha_0$ = (a) $0.225$, (b) $0.45$, (c) $0.225$, and (d) $[0.225,0.45]$, $k_{F0} =$ (a) $0.25$, (b-c) $[0.15,0.35]$, with $k_{g}=\alpha_{0}/25 - 0.0075$. Results presented are based on the average of 50 simulated colonies.}
\end{figure}

Now, consistent with the results of our hard-rod model presented in Sec. \ref{sec:mechanics}, we encode a specific density dependence in the quantities $p$, $\alpha$, and $k_{F}$. We introduce the packing fraction $\phi=\rho/\rho_{c}$, where $\rho_c$ is the density at which cells become closely packed and start to transmit stress. In the following, we assume $\rho_{c}>\rho^*$ to reflect the earlier observation that, at very low density (i.e., at the boundary of the colony), the contact forces tend to reorient cells rather than compress them. Based on these considerations, we set
\[
p = p_{0}\,(\phi-1)\;,\quad 
\alpha = - \alpha_{0}\,(\phi-1)\;,\quad 
k_{F} = k_{F0}\,(\phi-1)\;,
\]
where $p_{0}$, $\alpha_{0}$, and $k_{F0}$ are positive constants. Furthermore, we take $\alpha_{0} \sim k_{g}$ and keep $p_0$ constant, based on the results summarized in Fig. \ref{fig:mechanics}(c). Equations \eqref{eq:hydrodynamic} have been numerically solved using a finite difference approach on a $351 \times 351$ collocated grid. Figure \ref{fig:sshd} shows a typical configuration obtained for sufficiently large growth rates, in terms of the nematic director and order parameter [Fig. \ref{fig:sshd}(a) and \ref{fig:sshd}(b)] and velocity field [Fig. \ref{fig:sshd}(c) and \ref{fig:sshd}(d)]. As for our hard-rod models, these consist of an ensemble of randomly oriented nematic domains, whose characteristic area remains uniform in the bulk of the colony. In order to make a quantitative comparison between our discrete and continuous models, we reconstruct the geometrical properties of the microdomains based on the following criterion. Given the orientation $\theta=\arctan(Q_{xy}/Q_{xx})/2$ of the nematic director, we define $\Theta$ as the coarse-grained $\theta$ field in which all values are sorted into bins; e.g., $2(n-1)\pi/m \le \theta < 2n\pi/m \implies \Theta=(2n-1)\pi/m$ for $n=1,\,2,\,\ldots,\,m$ (with $n$ and $m$ both integers). This divides the colony into domains that can then be identified by labeling the connected components of the resulting two-dimensional matrix. We use a value of $m=6$ here to reflect a typical $\theta$ change between two boundaries in the hard-rod model.

Figure \ref{fig:resuhd} summarizes the results obtained from a numerical integration of Eqs. \eqref{eq:hydrodynamic}. As for the hard-rod model, the density decreases monotonically from the center of the colony, consistent with the quadratic law given by Eq. \eqref{eq:density_profile} [Fig. \ref{fig:resuhd}(a)]. Here, we demonstrate that such a property originates from the interplay between growth-induced pressure and drag from the substrate. Under this hypothesis and assuming low Reynolds number, from Eq. (\ref{eq:hydrodynamic}b), one can approximate the momentum density in the Darcy-like form $\rho \bm{v}=-\mu \nabla\rho$, where $\mu=p_{0}/(\xi \rho_{c})$ is a mobility coefficient. Using this relation in Eq. (\ref{eq:hydrodynamic}a) yields the following moving boundary value problem for the colony density:
\begin{subequations}\label{eq:stefan}
\begin{align}
&\partial_{t}\rho = \mu \nabla^{2}\rho + k_{g}\rho\;, \qquad |\bm{r}|<|\bm{R}|\;, \\[5pt]
&\rho(\bm{R},t) = \rho_{c}\;, \\[5pt]
&\dot{\bm{R}} = -\mu\rho^{-1}\nabla\rho|_{\bm{r}=\bm{R}}\;,
\end{align}
\end{subequations}
where we indicate with $\bm{R}$ the position of the boundary of the colony and with $\dot{\bm{R}}=\bm{v}(\bm{R})$ its velocity. Because of the circular symmetry of the colony at long times, $\bm{R}=R\bm{\hat{r}}$, and Eqs. \eqref{eq:stefan} reduce to a Stefan problem with one spatial and one temporal variable \cite{Ockendon:2003}. At short times, density and pressure are still roughly uniform across the system, and growth results mainly in a radial expanding flow. Consistent with Eq. (\ref{eq:hydrodynamic}a), if $\rho(r,t)\approx {\rm const}$, then $\nabla\cdot\bm{v}=k_{g}$. Thus, assuming $v_{\varphi}=0$, we get $v_{r}=k_{g}r/2$ and $R(t)=R(0)\exp(k_{g}t/2)$. The long time dynamics, on the other hand, is dominated by the internal diffusive currents. In this regime, $\rho(0,t)\gg \rho_{c}$ and $R \gg \sqrt{\mu/\kappa_{g}}$, which is the characteristic length scale associated with Eq. (\ref{eq:stefan}a). Thus, taking $\rho_{c} \rightarrow 0$ and $R\rightarrow\infty$, one can find an analytical solution of Eqs. \eqref{eq:stefan} of the form
\begin{equation} 
\rho(r,t) = \frac{M_{0}}{4\pi \mu t}\,\exp\left(k_{g}t-\frac{r^{2}}{4\mu t}\right)\;, 	
\end{equation}
under the assumption that $\rho(\bm{r},0)=M_{0}\delta(\bm{r})$. Thus, in agreement with Eq. \eqref{eq:density_profile}, we have
\begin{equation}
\frac{\rho(r,t)}{\rho(0,t)} \approx 1-\left(\frac{r}{R}\right)^{2}\;,	
\end{equation}
where, consistently with Eq. (\ref{eq:stefan}c), we have taken $R=2\sqrt{\mu t}$. For generic $\rho_{c}$ and $R$ values, Eqs. \eqref{eq:stefan} become analytically intractable; nonetheless, our numerical simulations [Fig. \ref{fig:resuhd}(a)] indicate that even the short time dynamics of the density $\rho$ is ultimately dominated by a similar competition between growth and drag. 

The geometrical properties of the nematic microdomains are summarized in Figs. \ref{fig:resuhd}(b)--\ref{fig:resuhd}(d). The area of the domains is exponentially distributed [Fig. \ref{fig:resuhd}(b)], and its average $\langle A \rangle_r$ is uniform across the colony [Fig. \ref{fig:resuhd}(c)] and proportional to the squared active length scale as demanded by Eq. \eqref{eq:average_area} [Fig. \ref{fig:resuhd}(d)]. The agreement between our discrete and continuous models not only validates our interpretation of the results presented in Sec. \ref{sec:md} but also demonstrates that the a growing bacterial colony can be described by the hydrodynamics theory of active nematics. On the one hand, this provides an efficient method to simulate growing bacterial colonies. Unlike discrete particle methods (including that used in Sec. \ref{sec:md}), our hydrodynamic approach does not suffer from the prohibitive slowdown caused by the exponential increase in the particle number, and it can be naturally generalized to other geometries and boundary conditions. On the other hand, this approach offers another prototype, i.e., growing bacterial colonies, for the experimental and theoretical study of active matter.

\section{\label{sec:exp} EXPERIMENT ON {\em E. COLI} MICROCOLONIES}
\subsection{\label{sec:meth}Experimental methods}

\begin{figure*}[t]
\centering
\includegraphics[width=\textwidth]{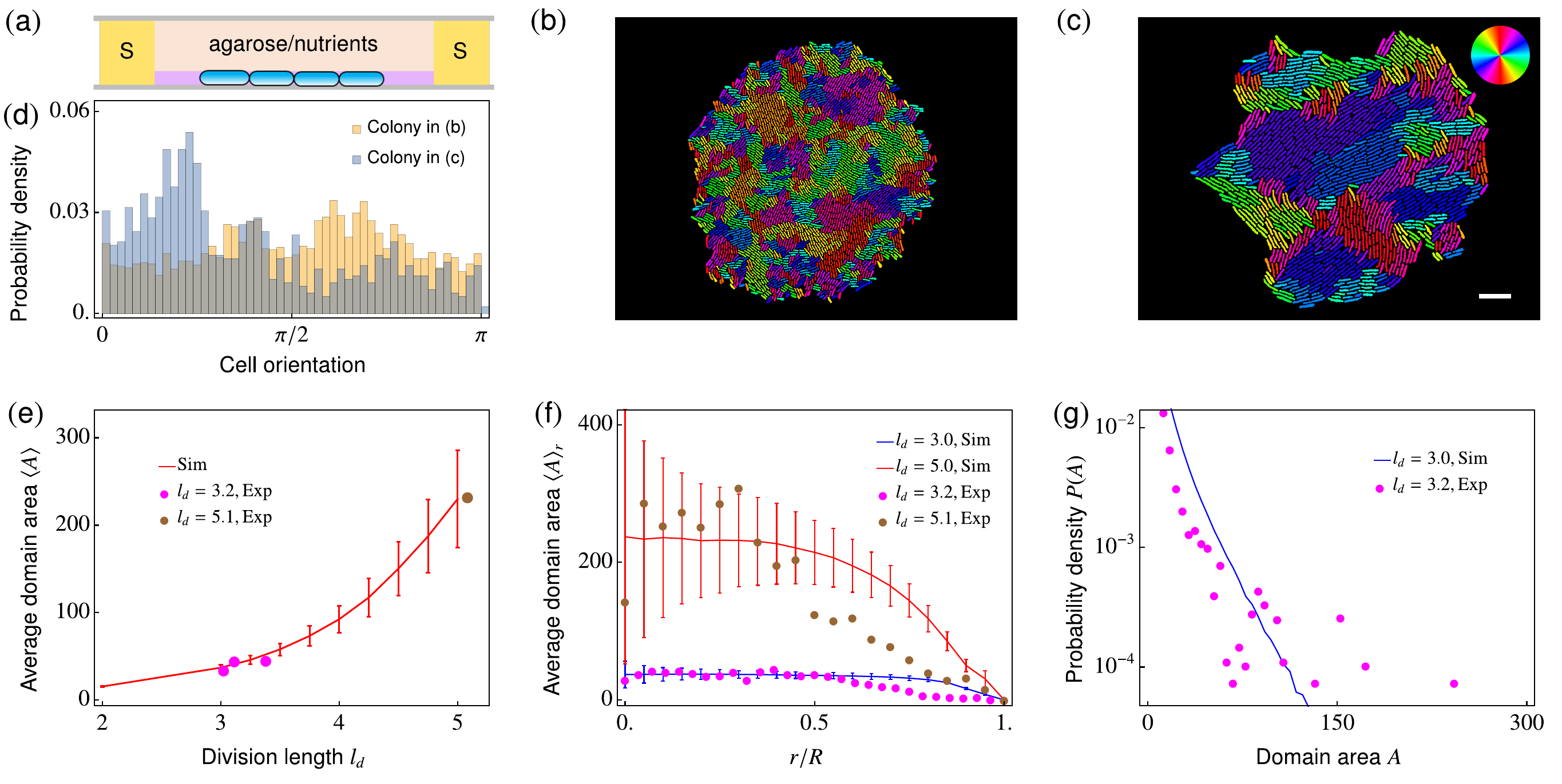}
\caption{\label{fig:exp} {\bf Experimental setup and results.} (a) Lateral view of the micro-environment used in the bacterial growth experiments. \textit{Escherichia coli} was grown on an agarose layer (pale pink) replete with Lysogeny broth (LB). The nutrient-rich agarose layer was sandwiched between two glass slides (grey hue) and enclosed with a 2-mm-thick neoprene spacer (marked as ``S'' in yellow). The cells (shown in blue) were imaged from below using time-lapse phase-contrast microscopy. (b,c) Snapshots of two-dimensional bacterial colonies with division lengths (b) $l_{d}=3.4$ and (c) $l_{d}=5.1$. Cells are color-coded with the same method as in Fig. \ref{fig:ssgrow}, and the scale bar corresponds to 10 $\mu$m. (d) Normalized frequency of orientation of cells shown in panels (b) and (c), shown in orange and blue, respectively. (e)--(g) Comparisons between the experiments and the simulations in (e) average domain area $\left\langle A \right\rangle$ as a function of division length $l_{d}$, (f) spatial distribution of domain area $\left\langle A \right\rangle_{r}$, and (g) probability density of domain area $P(A)$. Data from four independent colonies in the experiment are analyzed, each corresponding to a dot in panel (e). Magenta dots represent results from three experimental colonies ($l_{d}=3.4, 3.1, 3.0$ and $g=3.1\mu$m/h, $2.9 \mu$m/h, $2.8\mu$m/h in physical units) with an average division length $l_{d}=3.2$, and the brown dots are from one experimental colony with division length $l_{d}=5.1$ and growth rate $g=4.2\mu$m/h in physical units. The simulation data, represented by solid lines, are the same as those shown in Figs. \ref{fig:geometry}(a)--\ref{fig:geometry}c, i.e., with a growth rate $g=0.0002$ (or $4\mu$m/h in physical units). In panels (f) and (g), magenta dots show the results averaged over the first three colonies for better statistics. } 
\end{figure*}

To further test the significance of our results, we compare our theoretical predictions with experiments on a nonmotile strain of \textit{E coli} NCM 3722 delta-motA. The cell-to-colony growth was observed on a 2-mm-thick layer of agarose gel uniformly mixed with LB, a nutritionally rich medium commonly used for growing bacteria [Fig.~\ref{fig:exp}(a)]. The nutrient layer was sandwiched between two glass slides and enclosed within a 2-mm-thick neoprene spacer. The cells were imaged from below using time-lapse phase-contrast microscopy. For each experiment, we cultured the cells overnight in the LB medium. A dilute concentration of this culture was used to spot single bacterium on the agarose surface, which subsequently grew into colonies. For each experiment, \textit{E. coli} was cultured overnight in the LB medium at 25  $^{\circ}$C. A dilute concentration of this culture was then used to spot single bacterium on the agarose surface, which served as nucleating sites for subsequent colonies.

 Under the given experimental conditions, the average doubling time of bacteria was $43.5\pm 2.2$ minutes (doubling time for each replicate in minutes was 42.86, 45.89, 44.42, and 40.76). Cells in the colony were 0.9 $\pm$ 0.1 $\mu$m wide, while the average cell length varied among different colonies. The four replicates considered here were obtained under room-temperature conditions, which was stable at approximately 22$^{\circ}$ during the course of the measurements. The variability in the cell division lengths is frequently observed within colonies growing under similar conditions, potentially because of the inherent variability in the probability of growth itself (also known as phenotypic heterogeneity) \cite{Lianou:2013}. Statistics were measured over four independent colonies. The nutrient-rich agarose layer was thus sufficiently thicker than the bacterial monolayer ($\simeq$ 1 $\mu$m), which ensured constant availability of nutrients during the entire duration of the experiments.

We used time-lapse phase-contrast microscopy to visualize the growth of two-dimensional bacterial colonies [Fig.~\ref{fig:ssgrow}(a)--\ref{fig:ssgrow}(d)]. Images were acquired using an Andor iXon Ultra 897 camera (8 $\mu$m/px) coupled to an inverted microscope (Nikon TE2000) with a $40\times$ air objective (additional 1.5$\times$ magnification was used in some cases). This gave us a resolution of $0.2$ $\mu$m ($0.13$ $\mu$m with additional $1.5\times$ magnification). For a 4-$\mu$m-long cell, this resulted in a resolution of 20 pixels/cell. Using subpixel resolution (achieved by Gaussian interpolation), we could further improve this by a factor 2, which provided us with sufficient resolution to reliably detect and segment single cells. As checks, we analyzed the correct segmentation area over the entire segmented area (true positive rate) and, as a complementary parameter, looked at the false-positive rates. Prior to time-lapse image acquisition, we identified and recorded multiple spots on the agarose layer where single bacterium was present. The microscope was automated to scan these prerecorded coordinates and to acquire, every 3 minutes, the images of gradually growing bacterial colonies. By recording the phase-contrast images over hours, we acquired the necessary data for quantifying growing bacterial colonies. We analyzed the phase contrast images to extract the dimensions (length and width), position (centroid), and the orientation of each cell using intensity thresholding routines available through open source image analysis software ImageJ. Upon extraction of the cell dimensions, and the corresponding centroids and orientations, we generated orientation maps of the colony using MATLAB (MathWorks). 

\subsection{\label{sec:expres}Results}

Four independent colonies were cultured under the same experimental conditions as specified in the previous section. Despite their approximately similar doubling time, variance in their growth rates (rate of elongation) was quite significant, as was the variance in their division lengths. Figures \ref{fig:exp}(b) and \ref{fig:exp}(c) show two examples of proliferating colonies of cells, each with different division lengths and, hence, different cell aspect ratios. Like in the simulations, cells self-organize into nematic domains of different sizes and shapes, and the typical domain area increases with the cell aspect ratio. Along the colony boundary, the cells are preferentially oriented along the tangential direction, whereas in the bulk, the domains are isotropically oriented [Fig. \ref{fig:exp}(d)].

For the strain of bacteria we used, the division rate (or doubling time) is constant at a given temperature; hence, the growth rate (rate of elongation) is approximately proportional to the cell aspect ratio \cite{Sheats:2017}. It is thus difficult to vary the growth rate and the cell aspect ratio independently in our experiment, as done in the simulations. However, as we can see from Figs. \ref{fig:geometry}(c)--\ref{fig:geometry}(d), the variation of domain size is more sensitive to the division length $l_{d}$, if $l_{d}$ and $g$ are linearly related. For this reason, we compare the experimental results with those of a fixed growth rate from the MD simulations.

Figure \ref{fig:exp}(e) shows the average domain sizes of the four colonies (each represented by a dot) as a function of the division length $l_{d}$. We can see that the average domain sizes $\langle A\rangle$ in experiments fall well within the region predicted by our simulations. The spatial distribution of domain size, i.e., $\langle A\rangle_{r}$, is approximately constant for $l_{d}=3.2$ [magenta dots in Fig. \ref{fig:exp}(f)], and overlaps well with that for $l_{d}=3.0$ from the simulations. Here, $\langle A\rangle_{r}$ for $l_{d}=5.1$ [brown dots in Fig. \ref{fig:exp}(f)] is also in the expected region. Note that $\langle A\rangle_{r}$ drops at a smaller $r/R$ in the experiments. This is because the colony radius $R$ is smaller in the experiment, and the relative thickness of the boundary layer, which contains smaller domains, is larger. The probability densities $P(A)$ for $l_{d}=3.2$ also collapse with that for $l_{d}=3.0$ in simulations [Fig. \ref{fig:exp}(g)], although the area range is smaller than the simulated one, as a consequence of the limited statistics of our experiments.

Because of the limited statistics, our experimental results do not allows us to formulate conclusive statements. However, the quantitative agreement between the experiment and theory, is encouraging in suggesting that some of the geometrical and mechanical aspects of bacterial microcolonies can indeed be conceptualized in the framework of active liquid crystals.

\section{\label{sec:discussion}DISCUSSIONS AND CONCLUSIONS}

Sessile bacteria communities have the extraordinary ability to colonize a variety of surfaces, even in the presence of nonoptimal environmental conditions. Such a process typically starts from a few or even a single cell that elongates and eventually divides at a constant rate, and this gives rise to highly complex two-dimensional and three-dimensional structures consisting of tightly packed and partially ordered cells. Colonies originating from a single bacterium initially develop in the form of a flat and circularly symmetric monolayer and, after reaching a critical population, invade the three-dimensional space forming stacks of concentric disk-shaped layers \cite{Su:2012,Grant:2014}. While in the monolayer form, bacterial colonies exhibit prominent nematic order; however, this does not propagate across the colony, and it remains confined to a set of microscopic domains of coaligned cells. Using molecular dynamics simulations, continuous modeling and, to a limited extent, experiments on {\em E. coli} microcolonies, we have demonstrated that these domains originate from the interplay of two competing forces. On the one hand, the steric forces between neighboring cells favor alignment. On the other hand, the extensile active stresses due to growth tend to distort the system and disrupt the local orientational order. This results in an exponential distribution of the domain area, with a characteristic length scale $\ell_{\rm a}=\sqrt{k_{F}/|\alpha|}$, where $k_{F}$ is the orientational stiffness of the nematic domains and $\alpha$ the magnitude of the deviatoric active stress.

Our work generalizes and extends previous studies on the self-organization of sessile bacteria under confinement \cite{Volfson:2008,Boyer:2011}. In Ref. \cite{Volfson:2008}, Volfson {\em et al}. demonstrated that, when confined in a narrow channel, duplicating nonmotile bacteria tend to organize into colonies characterized by a strong orientational order. This effect was ascribed to the self-generated expansion flow induced by the bacterial growth. For large confinements, the ordering mechanism becomes less efficient, and the system transitions toward a disordered state consisting of multiple domains of aligned cells with no global order. A possible explanation of this transition was provided by Boyer {\em et al}. \cite{Boyer:2011} upon modeling the bacterial colony as an elastic continuum subject to an internal active pressure. According to this interpretation, the colony undergoes a buckling instability triggered by the growth-induced axial compression. The approach introduced here extends this by explicitly accounting for the internal nematic order and the hydrodynamic flow, thus broadening the scope for numerically investigating the post-transitional scenarios as well. In addition, the present work allows an accurate description of the disordered state with a number of experimentally testable predictions, such as the exponential distribution of the domain area, summarized by Eq. \eqref{eq:pdf}, and the dependence of the average domain area on the cell aspect ratio and growth rate [Figs. \ref{fig:geometry}(b) and \ref{fig:geometry}(c)]. The identification of the active length scale $\ell_{\rm a}$ offers a coherent interpretation of the collective behavior of confined and free-growing colonies alike. As active nematic liquid crystals, colonies on nonmotile duplicating bacteria are expected to be found in either an order or disordered state depending on the ratio between $\ell_{\rm a}$ and the system size $L$ (i.e., the confinement length scale in this case). When $\ell_{\rm a} \gg L$, the system relaxs toward orientationally ordered configurations, such as those discussed in Ref. \cite{Volfson:2008}, as the restoring forces arising in response to the elastic distortions outweigh the active forces. When $\ell_{\rm a} \ll L$, on the other hand, the system is spatially disordered and dynamically chaotic. Freely growing colonies, such as those studied here, correspond to the $L\rightarrow\infty$ limit and evolve directly into a disordered state characterized by a single length scale $\ell_{\rm a}$.

Even though cell morphology being one of the most well-documented phenotypic traits of microorganisms, its role as a functional trait in microbial ecology and evolution has received little attention \cite{Smith:2017}. The spontaneous creation of microdomains during the initial stages of colony growth presents a remarkable setting, one in which \textit{nonmotile} bacterial cells collectively lead to \textit{emergent motility} within the colony, as visualized in the chaotic fracture and coarsening dynamics of the nematic domains. Consequently, this interplay between growth-induced stresses and phenotypic stiffness of the participating cells introduces a novel angle to the transport and material attributes of such biologically active matter. Future studies on emergent motility within colonies of nonmotile cells, both in experiments and theory, are expected to contribute to a comprehensive biomechanical picture, highlighting the activity-driven cell-cell communications that precede biofilm formation. Finally, the results presented here are general and can be extended beyond bacterial communities, for instance, to study mammalian cells, many of which exist as nonmotile elongated phenotypes \cite{Duclos:2017}.

\acknowledgments

Z. Y., D. J. G. P, and L. G. are supported by The Netherlands Organization for Scientific Research (NWO/OCW) as part of the Frontiers of Nanoscience program. A.S. was supported by the International Human Frontier Science Program Organization (LT000993/2014-C) and the Vidi scheme. We are indebted to Jennifer Nguyen, Roman Stocker, and Cristina Marchetti for discussions.

\end{document}